\documentclass[a4paper]{article}
\usepackage[ansinew]{inputenc}
\usepackage{times}
\usepackage[T1]{fontenc}
\usepackage{graphicx}
\usepackage{geometry}
\usepackage{amssymb}
\usepackage{amsmath}

\usepackage{rotating}

\usepackage{tcolorbox}
\usepackage{xcolor}
\colorlet{shadecolor}{gray!25}
\usepackage{subfig}
\usepackage{float}

\author{\normalsize Ludwig A. Hothorn,\\ 
\footnotesize Im Grund 12, D-31867 Lauenau, Germany (e-mail:ludwig@hothorn.de)\\ \scriptsize(retired from Leibniz University Hannover)}

\title{The Kruskal Wallis test can not be recommended}
\begin{document}

\maketitle
\begin{abstract}
Although the Kruskal-Wallis (KW) test is widely used, it should not be recommended: it is not robust to arbitrary alternatives, it is only a global test without confidence intervals for the marginal hypotheses, it is inherently defined for two-sided hypotheses, it is not very suitable for pre/post hoc test combinations and hard to modified for factorial designs or the analysis of covariance. As an alternative a double maximum test is proposed: a maximum over multiple contrasts against the grand mean (approximating global power as a linear test statistics) and a maximum over three rank scores, sensitive for location, scale and shape effects. The joint distribution of this new test is achieved by the multiple marginal models approach. Related R-code is provided.
\end{abstract}


\section{The Kruskal Wallist test and three alternative proposals}
The Kruskal-Wallis test \cite{kruskal1952} is among the most widely used tests for analyzing randomized one-way designs  with treatment groups $T_j$:$[T_1, T_2, ..., T_k]$ as a nonparametric alternative to ANOVA F-test. Its test statistics represent a quadratic form for global ranks $R_{ji}$ (in $j$ groups with $i$ replicates) for a continuous variable: KW=$\sum_{j=1}^k n_j [\frac{ \sum_{i=1}^{n_j} R_{ji}}{n_j}-\frac{N+1}{2}]^2/(N(N-1)/12$ which is asymptotically $\chi^2$ distributed or related permutation version are available \cite{Hothorn2008a}. For tied data a modified variance estimator is available. The null hypothesis $H_0: F_1=F_2=...=F_k$ is against the alternative of any heterogeneity, at least $F_j\neq F_{j'}$, i.e. a global test only. However, in most k-sample layouts, the inference between the treatments is of interest, not just a global outcome of any heterogeneity. Its use as pre-test before post-hoc tests is a confusing concept using quite different alternatives conditionally \cite{LH2016}. It does not provide simultaneous confidence intervals for the interesting marginal hypotheses, is hard to generalize for factorial layouts or adjusting for covariates (ANCOVAR) and is inherently for two-sided hypotheses only.\\
Here, an alternative approach is proposed which based first on the similar power between ANOVA F-test and a multiple contrast test for comparisons the treatments against the grand mean $T..$ (MCT-GM) \cite{Konietschke2013}. Therefore, a non-parametric MCT-GM \cite{Pallmann2016} is considered, as a non-parametric global-rank test for relative effect size \cite{Konietschke2012} \cite{ksh2015} and as maxT-test over three scores tests (joint double maximum test (Joint)). The maxT-test based on the multiple marginal model approach (mmm) \cite{Pipper2012}: a first maximum over the multiple contrasts, a second maximum over rank-transformed responses (RT, sensitive against location effects), Ansari-Bradly scores (AB, sensitive against scale effects) and Savage scores (SA, sensitive for Lehman-type alternatives:
$T^{Joint}=max(F_{T_j-T..}^{RT},F_{T_j-T..}^{AB}, F_{T_j-T..}^{SA}$. The use of these three scores tests was motivated by a related sum-test over these scores tests \cite{mukherjee2022}. $T^{Joint}$ is $3k$-variate normal distributed with the correlation matrix $R$ between these $3k$ test statistics- estimated via mmm.

\section{Simulations}
In a tiny simulation study, size and power of the permutative Kruskal-Wallis test (KW), the joint test (Joint Test), the global rank nonparametric MCT (NonparMCT) and the MCT-GM for most likely transformations (MLT) \cite{Hothorn2018} are compared  in a balanced one-way layout with k=4, $n_i=20$ for Gaussian distribution and a skewed distribution in the Fleishman system (skewness=1.5, kurtosis=3) \cite{fleishman1978}. Three particular alternatives are considered i) just location, ii) just scale and iii) location-scale:

\begin{table}[ht]
\centering
\scalebox{0.85}{
\begin{tabular}{lllrrrr}
  \hline
Distribution & location & scale & Joint Test & NonparMCT & KW-test & MLT \\ 
  \hline
	
Normal & $H_0$ & $H_0$ & \textit{0.065} & 0.051 & 0.046 & 0.055 \\ 
Normal & $H_1$ & $H_0$ & \textbf{0.851} & 0.821 & 0.825 & 0.844 \\ 
Normal & $H_0$ & $H_1$ & \textbf{0.765} & 0.033 & 0.109 & 0.206 \\ 
Normal & $H_1$ & $H_1$ & \textbf{0.853} & 0.121 & 0.245 & 0.386 \\ \hline

Skewed & $H_0$ & $H_0$ & \textit{0.067} & 0.057 & 0.053 & 0.048  \\ 
Skewed & $H_1$ & $H_0$ & 0.910 & \textbf{0.975} & 0.908 & 0.888 \\ 
Skewed & $H_0$ & $H_1$ & \textbf{0.799} & 0.074 & 0.163 & 0.302 \\ 
Skewed & $H_1$ & $H_1$ & \textbf{0.975} & 0.913 & 0.647 & 0.543 \\ 
	
   \hline
\end{tabular}
}
\caption{Simulation results: size and power of 4 global tests }
\end{table}

Of course, there can be no most powerful test for these very different alternatives, but the joint test shows consistently high power in these simulations (in bold).

\section{An example}
As example the reaction time data of mice in a control and 3 treatments design is used \cite{Shirley1977}:

\begin{figure}[h]
	\centering
		\includegraphics[width=0.575\textwidth]{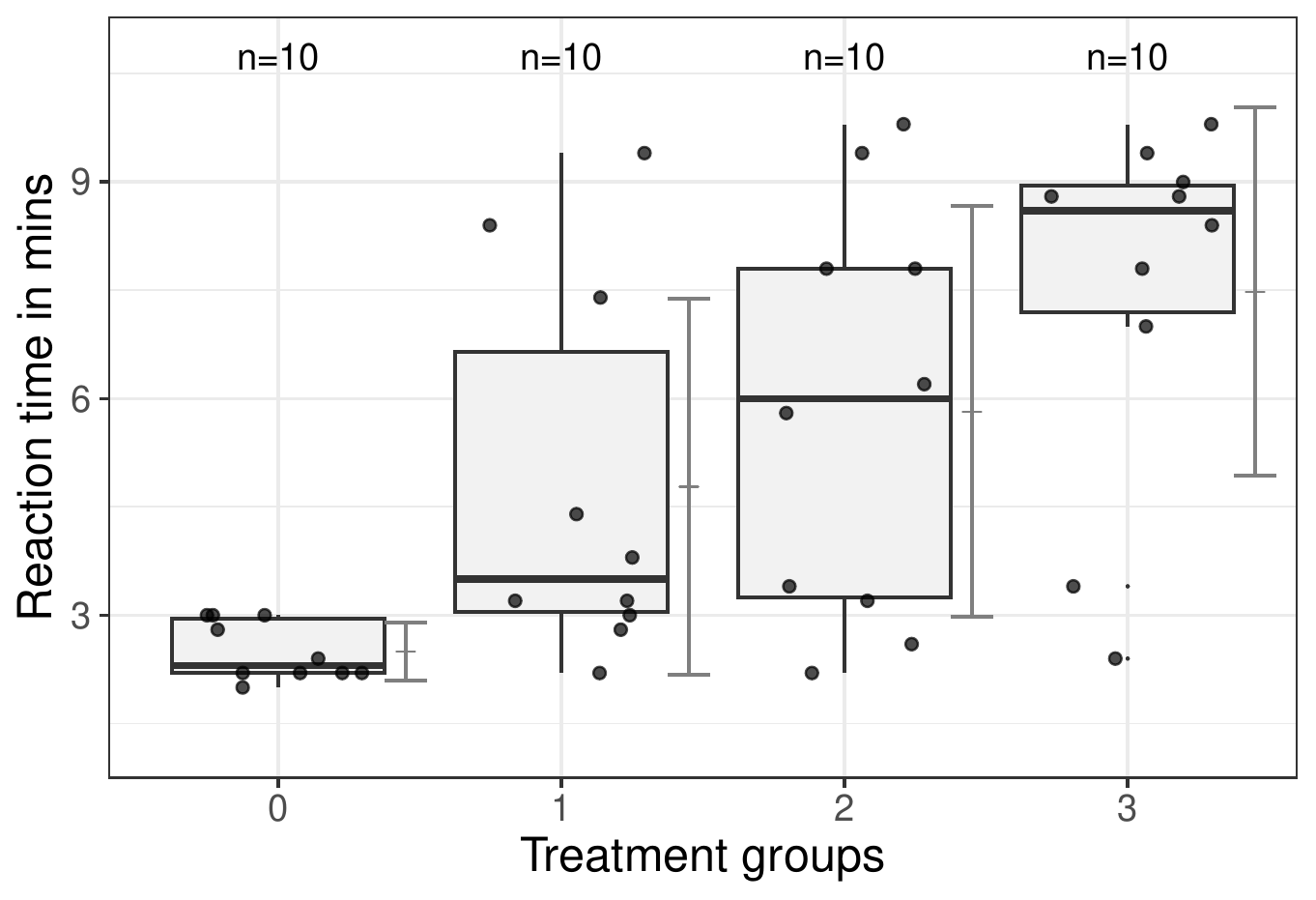}
	\caption{Shirley's reaction time data in mice}
	\label{fig:React}
\end{figure}

Because the design of this example is $[C, T_1, T_2, T_3]$, two types of analyses are compared: i) the global test version, ii) the one-sided Dunnett-type version for comparing the treatment groups against the control only.

\subsection{Global test}
The p-values of the Kruskal-Wallis test ($p=0.0007$), the MLT (p=0.0015), the NonparMCT-GM ($p<0.0001$) and Joint test ($p<0.0001$) are all small, but the joint test provides more detailed information:
\begin{table}[H]
\centering\footnotesize
\begin{tabular}{ll|rr}
  \hline
Effect& Treatment vs. GM & test stat & p-value \\ 
  \hline
location & 0 &  -5.90 & \textbf{0.00001} \\ 
  location & 1 & 0.07 & 1.00000 \\ 
  location & 2 & 1.04 & 0.90901 \\ 
  location & 3 &  2.99 & \textbf{0.04382} \\ \hline
  scale & 0 &  -1.51 & 0.65375 \\ 
  scale & 1 &  1.57 & 0.60633 \\ 
  scale & 2 &  0.61 & 0.99412 \\ 
  scale & 3 &  -0.81 & 0.97239 \\ \hline
  shape & 0 &-4.63 & \textbf{0.00061} \\ 
  shape & 1 & -0.40 & 0.99959 \\ 
  shape & 2 & 0.77 & 0.97975 \\ 
  shape & 3 &2.17 & 0.25429 \\ 
   \hline
\end{tabular}
\caption{Shirley example: adjusted p-values of the joint global test}
\end{table}

\subsection{Dunnett-type evaluation}

The one-sided Dunnett-type version for comparing the treatment groups against the control is for the above discussed test versions:

\begin{table}[H]
\centering\footnotesize
\begin{tabular}{ll|lll}
  \hline
Effect & Treatment vs. control & NonparamMCT & MLT & Joint \\ 
  \hline
  'location' &  1 - 0 & 0.0022   & 0.0141 &  0.01703 \\ 
  'location' &  2 - 0 & 0.0012 & 0.0048 &  0.00295 \\ 
  'location' &  3 - 0 & 0.000036 & 0.0004 & 0.00007 \\ 
  scale    & 1 - 0 & - & - & 0.19622 \\ 
  scale    &  2 - 0 & - & - & 0.46656 \\ 
  scale    &  3 - 0 & - & - & 0.90015 \\ 
  shape    &  1 - 0 & - & - & 0.25311 \\ 
  shape    &  2 - 0 & - & - & 0.04242 \\ 
  shape    &  3 - 0 & - & - & 0.00306 \\ 
   \hline
\end{tabular}
\caption{Shirley example: adjusted p-values 3 versions of Dunnett-type tests}

\end{table}

The experimental question is answered directly, that all groups cause an increase of the reaction time and the higher the dose the more significant - and this with very small p-values. The amount of location effect shows the plot with the simultaneous lower confidence limits, at the highest dose at least an increase of 10.6 min:
\begin{figure}[H]
	\centering
		\includegraphics[width=0.4\textwidth]{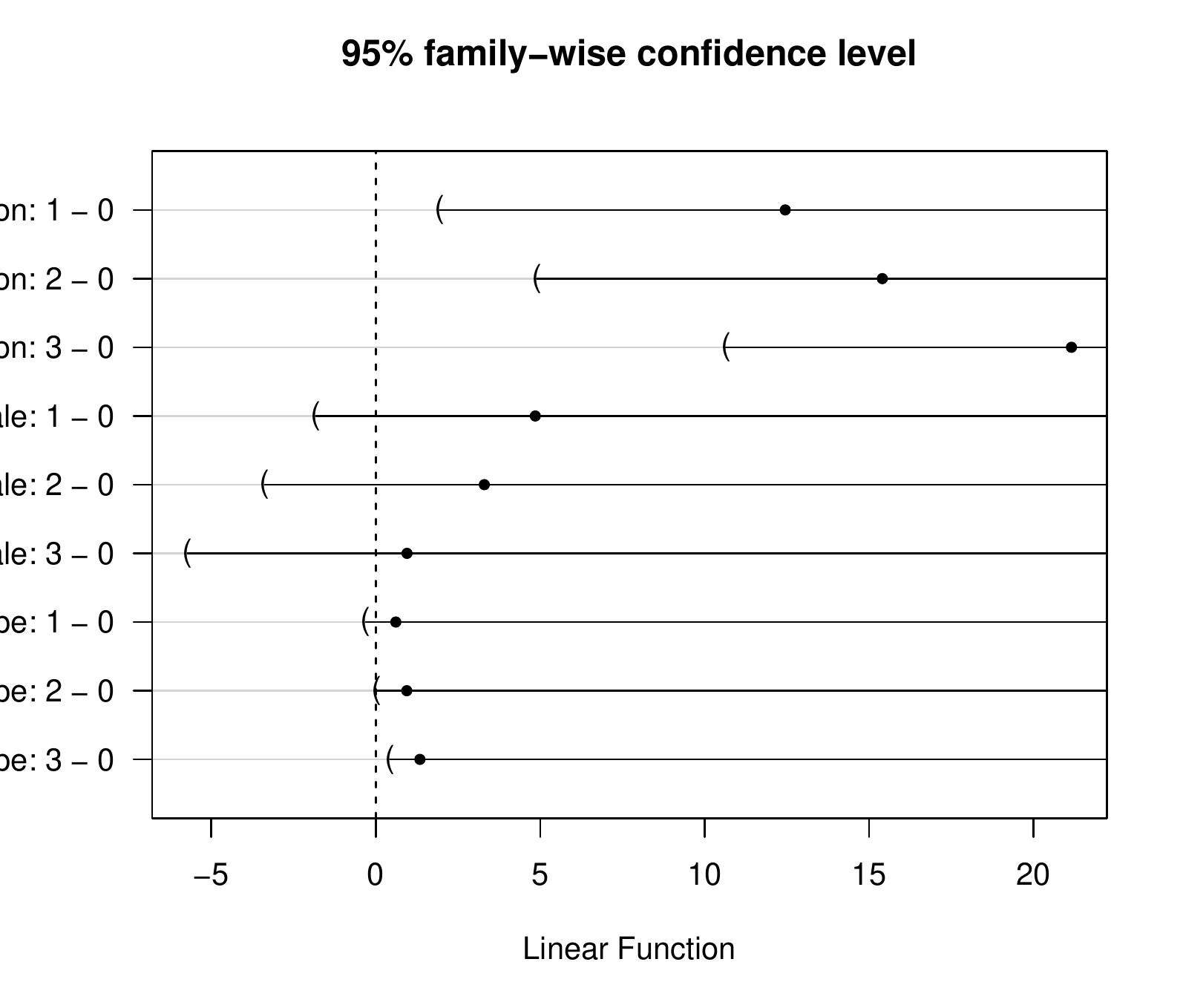}
	\caption{Shirley example: lower confidence limits}
	\label{fig:CIPLOT}
\end{figure}

\section{Conclusions}
It is clear that even for a global test for an one-way layout for arbitrary distributions no most powerful test can exist. 
The alternatives to the KW-test proposed here meet or exceed its power, but prepare additional information,  namely adjusted p-values and, or simultaneous confidence intervals on the marginal hypotheses and the joint test provides even information with respect to the underlying location/scale/shape effects. The related R-code is simple and provided within a data example.\\
Therefore, in summary, I suggest to use one of the alternative global tests instead of the KW-test, or even better instead of a global test (alone or within a pre-test/post-hoc test system) to use a test for the marginal hypotheses of interest, e.g. for comparisons with a control.

\section{R-Code}
\footnotesize
\begin{verbatim}
library(nparcomp)
data(reaction) 
library(toxbox)
boxclust(data=reaction, outcome="Time", treatment="Group",  ylabel="Reaction time in mins",  xlabel="Treatment groups", 
         option="uni", hjitter=0.3, legpos ="none", printN="FALSE", white=TRUE, psize=1.4, vlines="bg")
reaction$group<-as.factor(reaction$Group)
ni<-table(reaction$group); DF=sum(ni-4)
npc<-mctp(Time~group, data=reaction, type ="GrandMean", alternative ="two.sided",
          asy.method = "mult.t", plot.simci = FALSE, control = NULL, info = FALSE,
          correlation = FALSE)

library(coin)
kwp<-pvalue(kruskal_test(Time~group, data=reaction, distribution ="approximate"))

reaction$AB<- ansari_trafo(reaction$Time, ties.method ="mid-ranks")
reaction$SA<-savage_trafo(reaction$Time, ties.method ="mid-ranks")
reaction$Rldl<- rank(reaction$Time) 
mod2<-lm(Rldl~group, data=reaction)
mod3<-lm(AB~group, data=reaction)
mod4<-lm(SA~group, data=reaction)
library(multcomp)
Joint2 <- glht(mmm(location = mod2, scale= mod3, shape=mod4), mlf(mcp(group ="GrandMean")), df=DF)

library(tram)
TO<-glht(Colr(Time~group, data=reaction), linfct = mcp(group = "GrandMean"), df=DF)

\end{verbatim}
\footnotesize

\bibliographystyle{plain}





\begin{thebibliography}{10}

\bibitem{fleishman1978}
A.I. Fleishman.
\newblock A method for simulating non-normal distributions.
\newblock {\em Psychometrika}, 43(4):521--532, 1978.

\bibitem{LH2016}
L.A. Hothorn.
\newblock {The two-step approach-a significant ANOVA F-test before Dunnett's
  comparisons against a control-is not recommended}.
\newblock {\em {Comm.Stat. A}},
  {45}({11}):{3332--3343}, {2016}.

\bibitem{Hothorn2008a}
T.~Hothorn, K.~Hornik, M.~A.~V. van~de Wiel, and A.~Zeileis.
\newblock Implementing a class of permutation tests: The coin package.
\newblock {\em Journal of Statistical Software}, 28(8):8, November 2008.

\bibitem{Hothorn2018}
T.~Hothorn, L.~Most, and P.~Buhlmann.
\newblock Most likely transformations.
\newblock {\em Scandinavian Journal of Statistics}, 45(1):110--134, March 2018.

\bibitem{Konietschke2013}
F.~Konietschke, S.~Bosiger, E.~Brunner, and L.~A. Hothorn.
\newblock Are multiple contrast tests superior to the ANOVA?
\newblock {\em International Journal of Biostatistics}, 9(1):63--73, May 2013.

\bibitem{Konietschke2012}
F.~Konietschke and L.~A. Hothorn.
\newblock Rank-based multiple test procedures and simultaneous confidence
  intervals.
\newblock {\em Electronic Journal of Statistics}, 6:738--759, 2012.

\bibitem{ksh2015}
F. Konietschke, M., F. Schaarschmidt, and L.A. Hothorn.
\newblock {nparcomp: An R Software Package for Nonparametric Multiple
  Comparisons and Simultaneous Confidence Intervals}.
\newblock {\em {Journal of Statistical Software}}, {64}({9}), {2015}.

\bibitem{kruskal1952}
W.H. Kruskal and W.A. Wallis.
\newblock Use of ranks in one-criterion variance analysis.
\newblock {\em Journal of the American Statistical Association},
  47(260):583--621, 1952.

\bibitem{mukherjee2022}
A. Mukherjee, W. K{\"o}ssler, and M. Marozzi.
\newblock A distribution-free procedure for testing versatile alternative in
  medical multisample comparison studies.
\newblock {\em Statistics in Medicine},
41:2978-3002, 2022.

\bibitem{Pallmann2016}
P.~Pallmann and L.~A. Hothorn.
\newblock Analysis of means: a generalized approach using r.
\newblock {\em Journal of Applied Statistics}, 43(8):1541--1560, June 2016.

\bibitem{Pipper2012}
C.~B. Pipper, C.~Ritz, and H.~Bisgaard.
\newblock A versatile method for confirmatory evaluation of the effects of a
  covariate in multiple models.
\newblock {\em Journal of the Royal Statistical Society Series C-Applied
  Statistics}, 61:315--326, 2012.

\bibitem{Shirley1977}
E.~Shirley.
\newblock Nonparametric equivalent of Williams test for contrasting increasing
  dose levels of a treatment.
\newblock {\em Biometrics}, 33(2):386--389, 1977.

\end{thebibliography}

\end{document}